\documentclass[usenatbib]{mn2e}
\usepackage{graphicx}

\title[Detectability of Exoplanetary Transits]
{Detectability of Exoplanetary Transits from Radial Velocity Surveys}
\author[S. R. Kane]{Stephen R. Kane\\
Department of Astronomy, University of Florida, 211 Bryant Space Science
Center, Gainesville, FL 32611-2055, USA}

\begin{document}

\maketitle

\begin{abstract}

Of the known transiting extra-solar planets, a few have been detected
through photometric follow-up observations of radial velocity planets. Perhaps
the best known of these is the transiting exoplanet HD 209458b. For hot
Jupiters (periods less than $\sim 5$ days), the a priori information that 10\%
of these planets will transit their parent star due to the geometric transit
probability leads to an estimate of the expected transit yields from radial
velocity surveys. The radial velocity information can be used to construct an
effective photometric follow-up strategy which will provide optimal detection
of possible transits. Since the planet-harbouring stars are already known in
this case, one is only limited by the photometric precision achieveable by the
chosen telescope/instrument. The radial velocity modelling code presented here
automatically produces a transit ephemeris for each planet dataset fitted by the
program. Since the transit duration is brief compared with the fitted period,
we calculate the maximum window for obtaining photometric transit observations
after the radial velocity data have been obtained, generalising for eccentric
orbits. We discuss a typically employed survey strategy which may contribute to
a possible radial velocity bias against detection of the very hot Jupiters
which have dominated the transit discoveries. Finally, we describe how these
methods can be applied to current and future radial velocity surveys.

\end{abstract}

\begin{keywords}
stars: planetary systems -- methods: observational
\end{keywords}

\section{Introduction}

The detection of extra-solar planets in recent years has led to significant
advances and challenges in theories regarding planet formation. In particular,
planetary migration \citep{for06} and atmosphere models \citep*{bur06} have been
faced with intense revision in light of the gradually revealed distribution of
exoplanetary orbital parameters. The major contribution to revealing this
distribution has resulted from large-scale radial velocity surveys, such as
those being conducted by the California \& Carnegie Planet Search \citep{mar97}
and the High Accuracy Radial velocity Planet Searcher (HARPS) \citep{pep04}
teams. Moreover, the transit method is experiencing increased success through
surveys such as the Transatlantic Exoplanet Survey (TrES) \citep{odo06}, the XO
project \citep{mcc06}, the Hungarian Automated Telescope Network (HATNet)
\citep{bak07a}, and SuperWASP \citep{col07}. The improving detection rate from
transit surveys is due in no small part to the increasing understanding of
optimal photometric methods for wide-field detectors \citep{har04} and the
reduction of correlated (red) noise \citep*{pon06,tam05}.

Radial velocity measurements of an extra-solar planet provide such information
as a lower-limit on the planetary mass along with the orbital period and
eccentricity. However, if we are fortunate enough to view the orbit nearly
edge-on, then measurements of the planetary transit provide a wealth of
complimentary information which reveals the true mass, the radius, and hence the
mean density of the planet \citep{sea03}. The discovery of these transits has
thus far been mostly due to the ``brute force'' approach of wide-field searches,
such as those mentioned above. In at least four instances though, the planetary
transits were detected for extra-solar planets which were already known via
their radial velocity discoveries. These four planets are HD 209458b
\citep{cha00,hen00}, HD 149026b \citep{sat05}, HD 189733b \citep{bou05}, and
GJ 436b \citep{gil07}; the physical characteristics of which span the large
diversity seen amongst the short-period ($<$ 5--10 days) planets known as hot
Jupiters. Further attempts have been made to detect possible transits of certain
radial velocity planets \citep{lop06,sha06} which, though unsuccessful, have
placed useful constraints on orbital parameters. For those with the interest and
the resources to participate in follow-up observations, transit ephemerides for
the known radial velocity planets are provided by a useful online
resource\footnote{http://www.transitsearch.org/}.

Various current and future radial velocity surveys are predicted to uncover
a large number of exoplanets \citep*{kan07}. Optimal period analysis and
modelling of the data can be used to produce a predicted transit ephemeris.
Since the percentage of transiting planets amongst hot Jupiters is around
10\% based upon the geometric transit probability (assuming a random
distribution of line-of-sight orbital inclinations), it is expected that
photometric follow-up of the radial velocity planets will lead to a
significant amount of transiting planet discoveries. The error bars on the
predicted observing windows will of course increase with the time elapsed
since the last data were obtained. It is therefore important to develop an
efficient observing strategy for photometric follow-up at the predicted
times.

There is an observed difference between the period distributions of planets
discovered by the transit method and the radial velocity method
\citep*{gau05,but06}. Even though the transit method is extremely sensitive to
short period planets, the fact that the radial velocity method is also
increasingly sensitive to shorter period planets has led to speculation as to
how much of the observed difference is real and how much may be attributed to
radial velocity selection effects. It is possible that the observing strategy
of many radial velocity surveys may contribute to a bias against detecting
the very hot Jupiters found by the transit surveys.

This paper presents a summary of how the radial velocity information acquired
on an extra-solar planet may be used to construct a photometric follow-up
strategy which is optimised towards detecting planetary transits. Section 2
outlines the orbital parameters measurable from radial velocity data and how
that leads to a predicted epoch of a planetary transit. Section 3 explains
the photometric follow-up strategy, including the transit ephemeris, observing
window, and the best telescope/instrument to use for the observations. Section
4 describes a possible correlation of radial velocity observing strategy with
the dearth in very hot Jupiter discoveries from radial velocity surveys.
Section 5 compares the optimal strategy with other proposed methods and
dicusses the application of the techniques to various current and future
radial velocity surveys for extra-solar planets.

\section{Transit Predictions}

The radial velocity and transit methods provide complimentary information
regarding both the orbital parameters of the planet and the physical
characteristics of the planet itself. This section describes how, provided
the orbital parameters are determined with sufficient accuracy from fitting
the radial velocity data, the times of possible transit can be predicted.

\subsection{Complimentary Information}

\begin{figure}
  \includegraphics[angle=0,width=8.2cm]{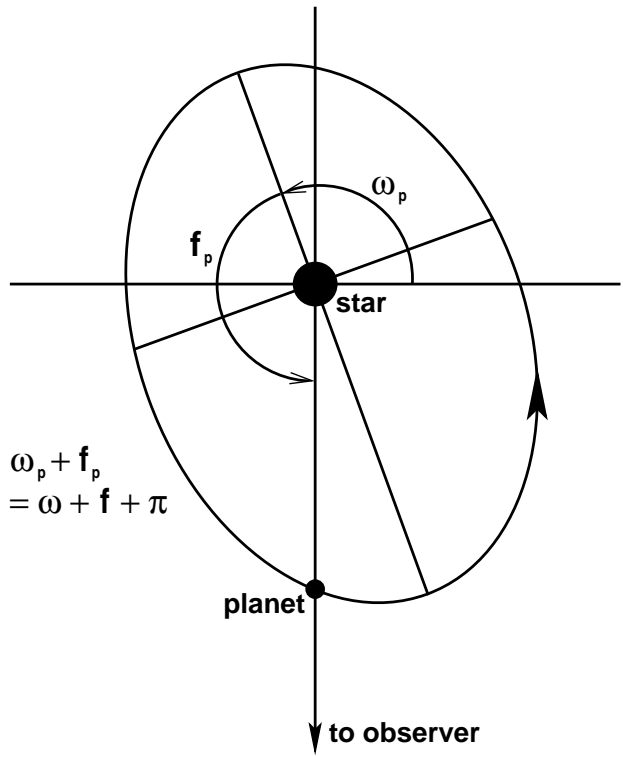}
  \caption{A planetary orbit as seen from above, showing the argument of
    periastron, $\omega_p$, and the true anomaly, $f_p$.}
\end{figure}

The information provided by radial velocity measurements allows one to
construct a model of the planetary orbit which only excludes the orbital
inclination. The measured radial velocity of the parent star, $V$, is given
by
\begin{equation}
  V = V_0 + K ( \cos(\omega + f) + e \cos \omega)
\end{equation}
where $V_0$ is the systemic velocity, $K$ is semi-amplitude, $\omega$ is the
argument of periastron, $f$ is the true anomaly, and $e$ is the eccentricity.
The argument of periastron is the angle between the plane of the sky and the
position at periastron. The true anomaly is the angle between the position at
periastron and the current position in the orbit measured at the focus of the
ellipse. The definitions of $\omega$ and $f$ for the planet (designated
$\omega_p$ and $f_p$) are demonstrated in Figure 1, where there is a $\pi$
phase shift relative to $\omega + f$.
The semi-amplitude of the radial velocity, $K$,
may be expressed as
\begin{equation}
  K = \left( \frac{2 \pi G}{P} \right)^{1/3}
  \frac{M_p \sin i}{M_t^{2/3}} \frac{1}{\sqrt{1-e^2}}
\end{equation}
where $P$ is the period, $i$ is the inclination of the planetary orbit, and
$M_t = M_p + M_\star$ is the total combined mass of the planet and parent star
respectively. The orbital period is also related to the semi-major axis of the
planetary orbit, $a$, via Kepler's third law:
\begin{equation}
  P^2 = \frac{4 \pi^2 a^3}{G M_t}
\end{equation}
Careful fitting of the data and spectral typing of the parent star will
therefore lead to estimations of $M_p \sin i$ along with $P$ and $e$.

Precision photometry obtained during a planetary transit uncovers further
information. In particular, the radius of the planet is extracted via the
flux difference inside and outside of transit:
\begin{equation}
  \frac{\Delta F}{F_0} = \frac{R_p}{R_\star}
\end{equation}
where $R_p$ and $R_\star$ are the radii of the planet and parent star
respectively. The complete light curve for the transit including the effect
of quadratic limb-darkening can be computed using the formalism shown in
\citet{man02}. The inclination of the orbit is readily resolved since the
transit duration, $t_d$, depends upon its magnitude:
\begin{equation}
  t_d = \frac{P}{\pi} \arcsin \left( \frac{\sqrt{(R_p+R_\star)^2 - (a \cos
      i)^2}}{a} \right)
\end{equation}
thus leading to a resolution of the $M_p \sin i$ ambiguity. Equation (5)
applies to the case of a circular orbit, but can be generalised to also
apply to eccentric orbits \citep{tin05}:
\begin{equation}
  t_d = 2 R_t \sqrt{1 - \frac{a_t^2 \cos^2 i}{R_t^2}}
  \frac{\sqrt{1 - e^2}}{1 + e \cos f} \left( \frac{P}{2 \pi G M_t}
  \right)^{1/3}
\end{equation}
where $R_t = R_p + R_\star$ is the total combined radii of the planet and
parent star. The dependence of the transit duration on the true anomaly in
this case leads to a shorter duration near periastron and a longer duration
near apastron.

The complimentary nature of the two detection techniques linked by their
common determination of the orbital period has lead to a successful
partnership of the two methods. In the context of this paper, the most
important aspect of the complimentary nature is the usefulness of the
constructed orbital model in the prediction of a possible planetary transit.
This will be described in detail in later sections.

\subsection{Transit Probability}

Geometrically, the probability of a planet passing between the observer and
the planet's parent star is naturally quite low. The probability of an
observable planetary transit occurring depends upon the inclination of the
planet's orbital plane $i$ satisfying $a \cos i \leq R_p + R_\star$. The
transit probability, $P_t$, is then given by
\begin{equation}
P_t = \frac{(R_p + R_\star)}{a (1 - e \cos E)} \approx \frac{R_\star}{a
  (1 - e \cos E)}
\end{equation}
where $E$ is the eccentric anomaly. For a circular orbit, this simplifies to
$P_t \approx R_\star / a$ from which it is clear that, as far as the probability
is concerned, the size of the planet is of little consequence and the dependence
lies mostly upon the size of the parent star and the orbital radius. However,
equations (4) and (7) show that the transit method clearly favours large planets
orbiting their parent stars at small orbital radii. Thus, many of the hot
Jupiters discovered via radial velocity surveys are expected to also exhibit a
photometric transit signature.

\begin{figure}
  \includegraphics[angle=270,width=8.2cm]{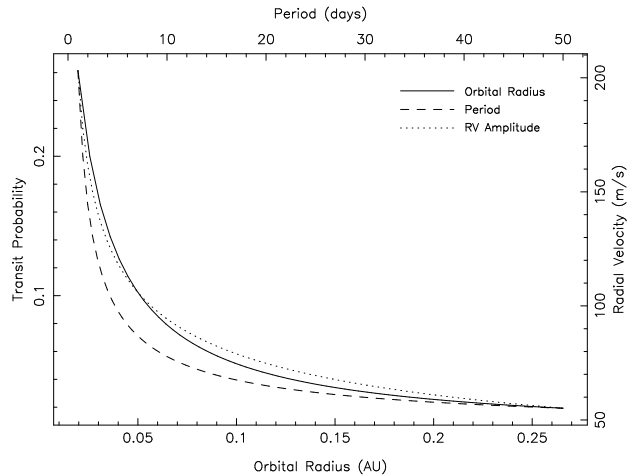}
  \caption{The probability of a Jupiter-radius planet transiting a solar-radius
    star for periods ranging between 1 day and 50 days. The peak radial velocity
    amplitude exhibited by such a star as a function of period is also plotted as
    a dotted line.}
\end{figure}

Figure 2 shows the dependence of the geometric transit probability on semi-major
axis and orbital period for a Jupiter-radius planet in a circular orbit around a
solar-radius star. The transit probability decreases very rapidly with increasing
period and leading to an almost negligible probability beyond a 50 day period
orbit. Thus, most of the currently known transiting extra-solar planets have
periods less than 5 days since this region of period space has by far the highest
probability of producing an observable transit. Figure 2 also shows the
equivalent peak radial velocity amplitude that would be produced by a transiting
Jupiter-mass planet orbiting a solar-mass star as a function of period, assuming
a circular orbit. This relation shows a similar dependence on period for the
given period range.

According to \citet{lin03}, 0.5\%--1\% of Sun-like stars in the solar
neighbourhood have been found to harbour a Jupiter-mass companion in a 0.05 AU
(3--5 day) orbit. Assuming that the orbital plane of these short-period planets
are randomly oriented, the geometric transit probability is approximately 10\%.
By performing Monte-Carlo simulations such as those conducted by \citet{kan07},
one can use this probability to predict the number of observable transits in
a given survey field. If the planets are first discovered by the radial
velocity method, then one is mostly able to avoid the false-alarm contaminants,
such as those described by \citet{bro03}, that plague purely photometric
transit surveys.

\subsection{Epoch of Planetary Transit}

\begin{figure*}
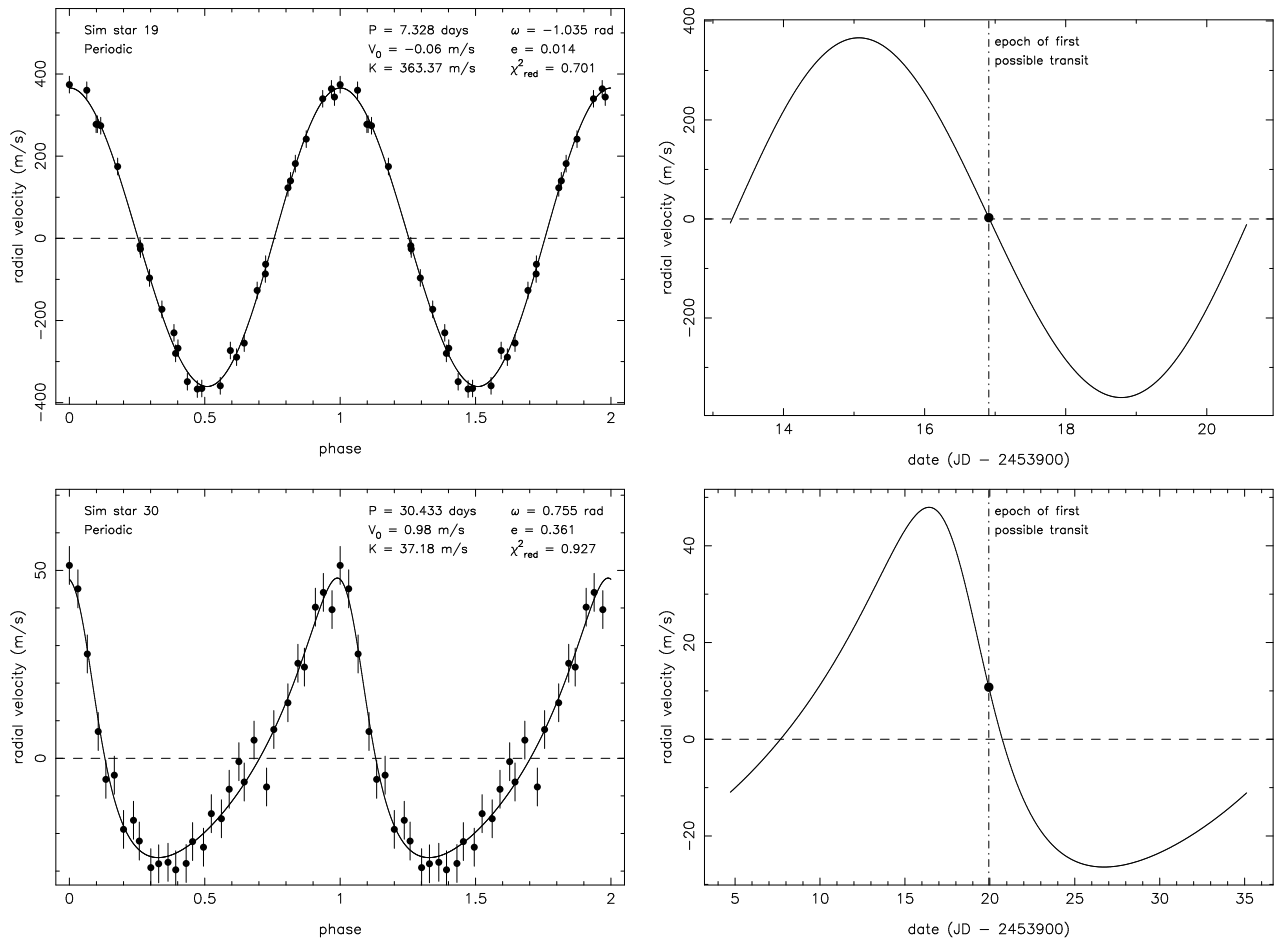

  \begin{center}
    \begin{tabular}{cc}
      \includegraphics[angle=270,width=8.2cm]{figure03a.ps} &
      \includegraphics[angle=270,width=8.2cm]{figure03b.ps} \\
      \includegraphics[angle=270,width=8.2cm]{figure03c.ps} &
      \includegraphics[angle=270,width=8.2cm]{figure03d.ps} \\
    \end{tabular}
  \end{center}
  \caption{The best-fit radial velocity model for two simulated datasets (left),
    each with an accompanying plot showing the best-fit model with the epoch of
    a possible transit time (right). The top example is for a short-period
    circular orbit and the bottom example is for a longer period elliptical
    orbit.}
\end{figure*}

Once a suitable fit is found for the radial velocity data, one can calculate
the epoch at which the next planetary transit may occur. For a purely circular
orbit, one expects that the transit will occur when the radial velocity of the
star is zero (peak-to-peak mid-point) and decreasing. An approximation of this
sort will suffice in most cases since the effects of tidal circularization are
well observed on hot Jupiters to produce close to zero eccentricity orbits. A
notable exception to this has been observed with the high eccentricity of the
transiting planet HD 147506b \citep{bak07b}. As described for binary systems by
\citet*{lec76} and compared to exoplanetary systems by \citet*{hal05}, the tidal
torque varies as $1/a^6$ leading to a quite severe dependence on period. Beyond
the tidal circularization limit of $\sim 0.1$~AU however, the range of orbital
eccentricities becomes extremely broad with little or no dependency on period.
A more general description of when a possible transit will occur is defined
using the argument of periastron and the true anomaly, as shown in Figure 1. In
other words, it is the location in the orbit where $\omega + f = \pi / 2$.
This is true regardless of if the orbit is prograde or retrograde. Minimising
the radial velocity function at this location then yields the time of possible
transit closest to the times of radial velocity observations.

According to equation (1), the difference in radial velocity between an
eccentric and circular orbit at the point of mid-transit is
\begin{equation}
  \Delta V = V_e - V_c = K e \cos \omega
\end{equation}
which can lead to a significant change in the predicted epoch of transit.
Moreover, the difference in the time of mid-transit can be calculated by
considering the mean anomaly, $M$, given by
\begin{equation}
  M = \frac{2 \pi}{P} (t - t_0) = E - e \sin E
\end{equation}
where $t_0$ is the time at periastron. The eccentric anomaly is related to the
true anomaly by
\begin{equation}
  \cos f = \frac{\cos E - e}{1 - e \cos E} = \cos (\pi / 2 - \omega)
\end{equation}
and allows equation (1) to be evaluated as a function of time. Hence, the
difference in time of mid-transit between an eccentric and circular orbit is
\begin{equation}
  \Delta t = t_e - t_c = - \frac{P}{2 \pi} (e \sin E)
\end{equation}
where one can evaluate $E$ by solving equation (9), also known as Kepler's
equation.

Figure 3 shows two examples of simulated radial velocity datasets, each with a
different planetary signature injected. The top example demonstrates a
relatively short-period planet with a typically circular orbit. In this case
the predicted epoch of transit occurs exactly when the decreasing radial
velocity reaches zero. The bottom example in Figure 3 demonstrates a longer
period planet with a substantially eccentric orbit in which the actual transit
time falls well outside the observing window of when a circular assumption
would have predicted. The magnitude of this time difference depends upon the
period, but the predicted time of transit can be in error by more than a day
for even a relatively short period. Since longer period planets have a much
smaller probability of producing an observable transit, these situations will
be relatively uncommon. However, considering searches in this region of
parameter space have already been conducted and future surveys will
undoubtedly be further exploring this region, the eccentricity of the orbit
must be taken into account.

\section{Photometric Follow-up Strategy}

This section describes how the radial velocity data and the predicted transit
times can be used to construct a complete follow-up observing strategy which
is optimised for the particular target star.

\subsection{Transit Ephemeris}

\citet{kan07} describes in detail a radial velocity simulation code called
{\tt rvsim}. One of the primary functions of this code is to ingest large
amounts of precision radial velocity data which can then be automatically
sifted for planetary signatures. The code also produces fits to the planetary
signals and provides transit ephemerides based on the methods presented in
the previous section. This is performed by first calculating the radial
velocity which satisfies $\omega + f = \pi / 2$. A grid search is then
used to locate the approximate time at which this occurs after the first
radial velocity measurement was obtained. Since this is a computationally
inexpensive algorithm, the grid precision can be exceptionally high without
loss of performance leading to an accurate estimate of predicted transit
epoch based on the fit parameters.

\begin{figure}
  \includegraphics[angle=270,width=8.2cm]{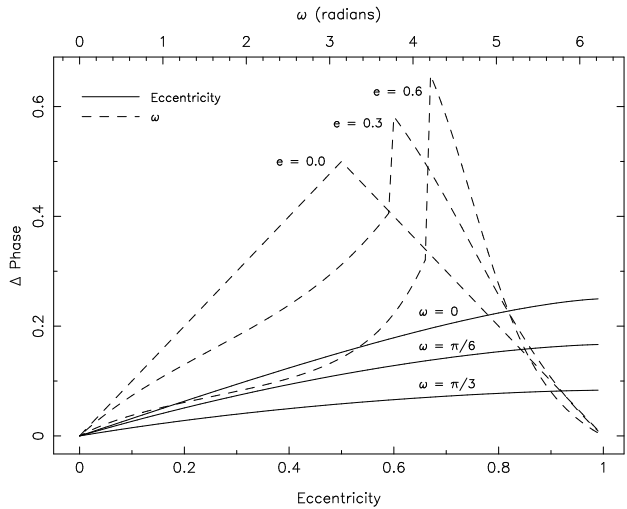}
  \caption{The effect of $\omega$ and $e$ on the predicted transit time,
    represented here as a fraction of the phase of the orbit. The solid lines
    shows the effect of $e$ for three different values of $\omega$ and the
    dashed lines show the effect of $\omega$ for three different values of
    $e$.}
\end{figure}

It has been already mentioned that the eccentricity of the orbit can have
a significant effect on the time of transit. It is also clear from the
definition of when an observable transit is predicted to occur and also from
Figure 1 that the ephemeris calculation is heavily dependent upon the
estimate of $\omega$. These two parameters tend to have a high degree of
uncertainty when relatively few data points are available to produce a fit.
The amplitude of the effect of the predicted transit time is demonstrated in
Figure 4 in which the change in transit time (represented as a fractional
amount of the phase or period) is plotted as a function of eccentricity and
$\omega$. Clearly these effects can become significant for even relatively
small uncertainties. For example, of the 172 planets tabulated in
\citet{but06}, 124 have $\omega$ uncertainties shown. The mean value of these
uncertainties is 20\degr and the median is 10\degr. For a circular orbit and
a period of 4 days, a shift of 10\degr \ in the value of $\omega$ will lead
to a 3 hour difference in the predicted time of mid-transit. Similarly, for a
4 day orbit with $e = 0.3$, a shift of 10\degr \ in the value of $\omega$
leads to a 2.5 hour difference. Constraining the values of $e$ and $\omega$
is therefore of critical importance before constructing a reliable transit
ephemeris.

As pointed out by \citet{wit05}, even very precise measurements of the orbital
period contain errors which propagate into larger uncertainties in the
ephemeris. This can take place in a relatively short period of time; a period
error of $\sim 1$ second will lead to an ephemeris uncertainty of several
minutes for hot and very hot Jupiters after only one year. Thus, though the
uncertainty in $e$ and $\omega$ can cause problems in predicting the epoch of
first transit, the error associated with the period has a cumulative effect on
the transit ephemeris. In the case of HD 189733b, we are fortunate that
detectable transits exist in the Hipparcos data which have allowed the period
to be determined with an accuracy of less than 1 second \citep{heb06}.
Maintaining an accurate ephemeris is important not only for future observations,
but for the possible detection of additional planets through transit timing
observations \citep{ago05}. Fortunately, ``re-alignment'' of the transit
ephemeris can always be re-established through further radial velocity
observations to re-predict the next transit event and further refine the
period.

\subsection{Transit Duration and Observing Window}

The predicted point of mid-transit at a particular epoch is only as accurate as
the robustness of the fit to the radial velocity data, in particular the period,
and the elapsed time since the most recent radial velocity measurements were
obtained. The size of the observing window to ensure complete coverage of the
transit can be estimated based on this information. Assuming that the values of
$e$ and $\omega$ are well constrained, as described in the previous section, the
observing window is defined by (a) the transit duration, (b) the uncertainty in
the measured period, and (c) the number of periods elapsed since the predicted
epoch of first transit. The dominant factor out of these is normally the transit
duration, which is given by equation (6). Figure 5 shows the variation in
observing window with period, including the effects of period uncertainty
($3\sigma$ either side of the transit duration) for an elapsed time of 20 orbits
and 50 orbits. This example assumes a Jupiter-radius planet in a circular orbit
transiting a solar-radius star with a period uncertainty of $\sim 5$ seconds.
The effect of eccentricity can be to increase or decrease the transit duration
by as much as 50\% up to an eccentricity of $\sim 0.5$, depending on the true
anomaly at the time of transit. This necessitates a proportional change in the
size of the observing window.

Even though the period of the planetary orbit can be quite accurately
determined, the transit duration will have a much larger associated uncertainty.
It was shown by \citet{kan05} that the duration of an exoplanetary transit is
relatively insensitive to the spectral type of the parent star as compared to
the effect of increasing orbital radius, and so the observing window shown in
Figure 5 is fairly representative. However, it is clear from equation (6) that
the estimated transit duration for a given orbital radius can depend heavily upon
the stellar radius estimate deduced from the spectral type. An example of this
can be seen in the detections of WASP-1b and WASP-2b \citep{col07} where the
difficulty presented by spectral typing imposed large uncertainties on the
stellar radius estimates. As such, the observing window should be calculated
based upon the assumption that the maximum estimates stellar radius and hence the
maximum estimated transit duration applies.

\begin{figure}
  \includegraphics[angle=270,width=8.2cm]{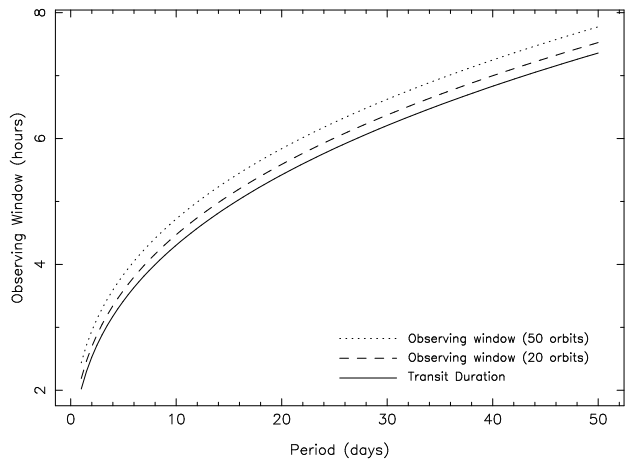}
  \caption{The observing window for detecting the transit of a Jupiter-radius
    planet orbiting a solar-radius star for periods ranging between 1 day and
    50 days. The solid line shows the transit duration. The dashed and dotted
    lines include the additional effect of the period uncertainty for 20 and
    50 orbits respectively after the last measurements were acquired.}
\end{figure}

\subsection{Rossiter-McLaughlin effect}

The Rossiter-McLaughlin (RM) effect \citep{mcl24,ros24} is a well-known
phenomenon observed for many eclipsing binary star systems, and has also been
predicted and observed in the radial velocity data acquired during exoplanetary
transits \citep*{oht05}. Furthermore, the confirmation of transiting planets and
measurement of planetary orbital alignments using the RM effect has been
discussed in detail by \citet{gau07}. Since a planetary transit is able to
manifest its presence in the radial velocity data, the RM effect allows the
remarkable opportunity to discover transiting planets through spectroscopic
means, as was the case for HD 189733 \citep{bou05}.

With respect to the impact the RM effect will have on the predicted transit
ephemeris, this depends upon the model used for the radial velocity data and
the percentage of data points which are obtained inside of the transit. As
described in \citet{oht05}, it is relatively trivial to include the RM effect
in the model being used to fit the radial velocity data. On the other hand, for
hot Jupiters the transit duration is typically 3--4 hours which is 3--5\% of the
total orbital period. Since there will be a correspondingly small number of
RM effect measurements in the complete dataset, it is possible to exclude these
measurements from the fit without compromising the orbital parameters extracted
from the model fit. This is the approach which was taken by \citet{sat05} who
also noted the small probability that they should acquire radial velocity data
during the transit of the hot Saturn orbiting HD 149026.

To determine the effect the RM effect has on the transit ephemeris, we produced
an independent fit to the radial velocity data published by \citet{sat05}. These
data contain 4 measurements during a transit, a remarkably large fraction of the
11 total data points. Including these measurements in the fit yields a period of
$P = 2.8767 \pm 0.002 \ \mathrm{days}$ and a semi-amplitude of $K = 43.1 \pm 1.7
\ \mathrm{m \ s^{-1}}$. Computing the transit ephemeris from this fit produced
predicted mid-transit times of $t_1 = 2453504.869$, $t_2 = 2453527.883$, and
$t_3 = 2453530.759$. When compared to the actual times of mid-transit measured
by \citet{sat05}, the differences are $\Delta t_1 = 0.004 \ \mathrm{days}$,
$\Delta t_2 = 0.019 \ \mathrm{days}$, and $\Delta t_3 = 0.008 \ \mathrm{days}$.
This is equivalent to differences of between 5 and 30 minutes. These differences
can be completely accounted for by the respective fit parameters obtained and the
cumulative stacking of the period uncertainty. Therefore in this case the RM has
had little effect on the transit ephemeris although a larger amplitude RM effect
with a comparable number of data points could prove fatal for transit predictions
attempted significantly after the radial velocity data is obtained. In such cases
an iterative elimination of radial velocity measurements within the transit
window (sigma-clipping) can enable a more accurate prediction of the transit
ephemeris.

\subsection{Transit Detectability}

The overwhelming majority of known radial velocity planets have been detected
as companions to stars with a magnitude of $V < 14$. Given the large number
of wide-field transiting planet surveys being conducted, it is possible that a
newly detected radial velocity planet will have already been monitored by one
or more of the surveys. However, large areas of the sky remain to be monitored
with significant time resolution and the phase coverage of transits that are
present in the data are likely to require follow-up observations. Therefore,
the optimal approach is to design a custom observing program which takes into
account the magnitude of the star and the required signal-to-noise (S/N) for
an unambiguous detection.

To demonstrate this, we design a simple example using a generic telescope and
detector. The telescope in this example is assumed to have a 1.0m aperture,
representative of the many under-subscribed 1.0m class telescopes which are
generally excellent choices for these observations. A typical detector would
have characteristics similar to a Tek 2K CCD, which for this example is
assigned a gain and readout noise of 2.4 e$^-$/ADU and 3.9 ADU respectively.
It is also assumed that a standard Cousins $R$ filter with a bandwidth of
$\sim 1500$ \ \AA is used with a total quantum efficiency of 50\%. The noise
model used takes into account detector characteristics as well as photon
statistics and takes the form
\begin{equation}
  \sigma^2 = \sigma_0^2 + \frac{(f_\star + f_\mathrm{sky}) \Delta t}{g}
\end{equation}
where $\sigma_0$ and $g$ are the CCD readout noise (ADU) and gain (e$^-$/ADU)
respectively, $f_\star$ and $f_\mathrm{sky}$ are the star and sky fluxes
respectively, and $\Delta t$ is the exposure time. For this simulation, we
adopt an exposure time of 30 seconds under grey time conditions.

The photometric accuracy will be limited by the number of suitable comparison
stars available within the instrument field-of-view (FOV). For a detector such
as the one used in this example, the typical pixel scale is 0.4 arcsec/pixel
giving a total FOV of $\sim 186$ square arcmins. Using the star counts for the
Kepler field presented by \citet{kan07}, the number of stars per square degree
is $\sim 96$ for $V < 12$ and $\sim 247$ for $V < 13$. This leads to an
estimate of $\sim 5$ stars in the FOV for $V < 12$ and $\sim 12$ stars in the
FOV for $V < 13$. Even for relatively bright targets, the integration times
can be adjusted to allow for sufficient differential photometry comparison
stars for in most cases. If this is not possible, then carefully considering
an instrument with a larger FOV may be required.

\begin{figure}
  \includegraphics[angle=270,width=8.2cm]{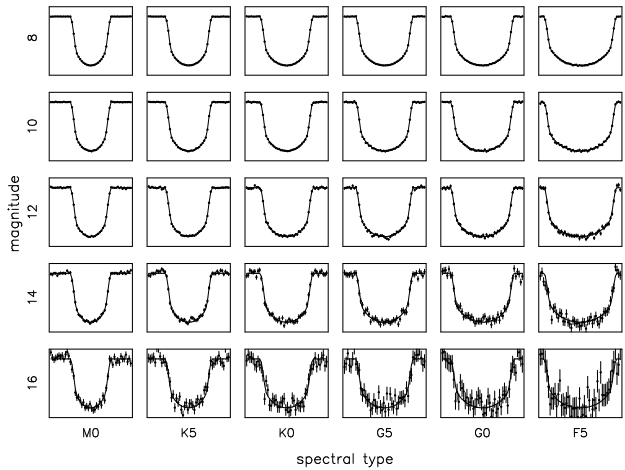}
  \caption{Model transit lightcurves overlaid with simulated data from
    a generic 1.0m telescope and detector for a range of spectral types and
    magnitudes. This assumes a Jupiter-radius planet transiting a
    solar-radius star.}
\end{figure}

Shown in Figure 6 are model transits and simulated data for a range of
magnitudes and spectral types assuming a single transit by a Jupiter-radius
planet. The plot windows have been normalised to a width of $3 R_\odot /
R_\star$, equivalent to the projected path of the planet as it crosses the
stellar disk, and a normalised depth for ease of comparison of photometric
accuracy. Transits of late-type stars, particularly M stars, are of shorter
duration but are considerably deeper and so produce a much higher S/N during
the transit. However, radial velocity surveys typically concentrate on F--G--K
stars for their brightness, frequency, and stability properties. This
simulation shows that conducting a suitable follow-up observing program with
reasonably accessable 1.0m class telescopes can lead to rapid confirmation or
elimination of observable planetary transits down to a magnitude depth which
is beyond most radial velocity surveys. 

The issues of red noise \citep{pon06} and stellar micro-variability
\citep*{aig04} have presented serious challenges to transit surveys. Various
tools have been developed to overcome these obstacles, such as the SysRem
algorithm developed by \citet{tam05} which is now commonly used to minimise
systematic effects in the photometry. The major way in which these problems
affect transit surveys is by generating false alarms of a quantity which is at
least an order of magnitude greater than the expected number of real transit
signatures. This becomes a problem that must be solved by an efficient transit
detection algorithm and a reliable set of rejection criteria. These
photometric calibration effects will also be present in the follow-up
photometric data of known radial velocity planets. These data have two key
advantages however. Firstly, many of the systematic noise properties of
transit surveys are due to the wide-fields, such as the position-dependent
airmass, colour, and point-spread function (PSF). These are present to a much
lesser extent in instrument designs such as the one presented here. Also,
since one only needs to observe during the predicted observing window, these
observations suffer far less from night-to-night variations which contribute
greatly to the red noise of transit surveys. Secondly, searching for transits
in the photometry of a single star is a very different problem than deriving
an automated method for scanning many thousands of stars. The targeted nature
of single-star follow-up allows for a much more thourough search of the
photometry to be conducted.

\section{The Radial Velocity Bias}

The currently known period distribution of extra-solar planets exhibits a
pileup of planetary periods which occurs near 3 days. This has been clearly
demonstrated by, for example, \citet{but06}. The mechanism of planetary
migration \citep{ric05} plays a major role in explaining much of the period
distribution observed for hot Jupiters. However, it has also been observed
that, although the radial velocity method is more sensitive to planets at
small orbital radii, the transit method has managed to uncover very hot
Jupiters (less than 3 day periods) which remained hidden to the radial
velocity surveys. It has been a matter for debate whether this lack of
radial velocity very hot Jupiters is an observational bias or selection
effect or if it is indeed a reflection of the real period distribution
\citep{gau05}. In other words, the transit detection of very hot Jupiters
may simply be due to the relatively high probability of detecting planets
at very small orbital radii even though such planets are quite rare. In this
section I discuss some of the radial velocity biases against detection of
planetary periods close to 1 day.

\begin{figure*}
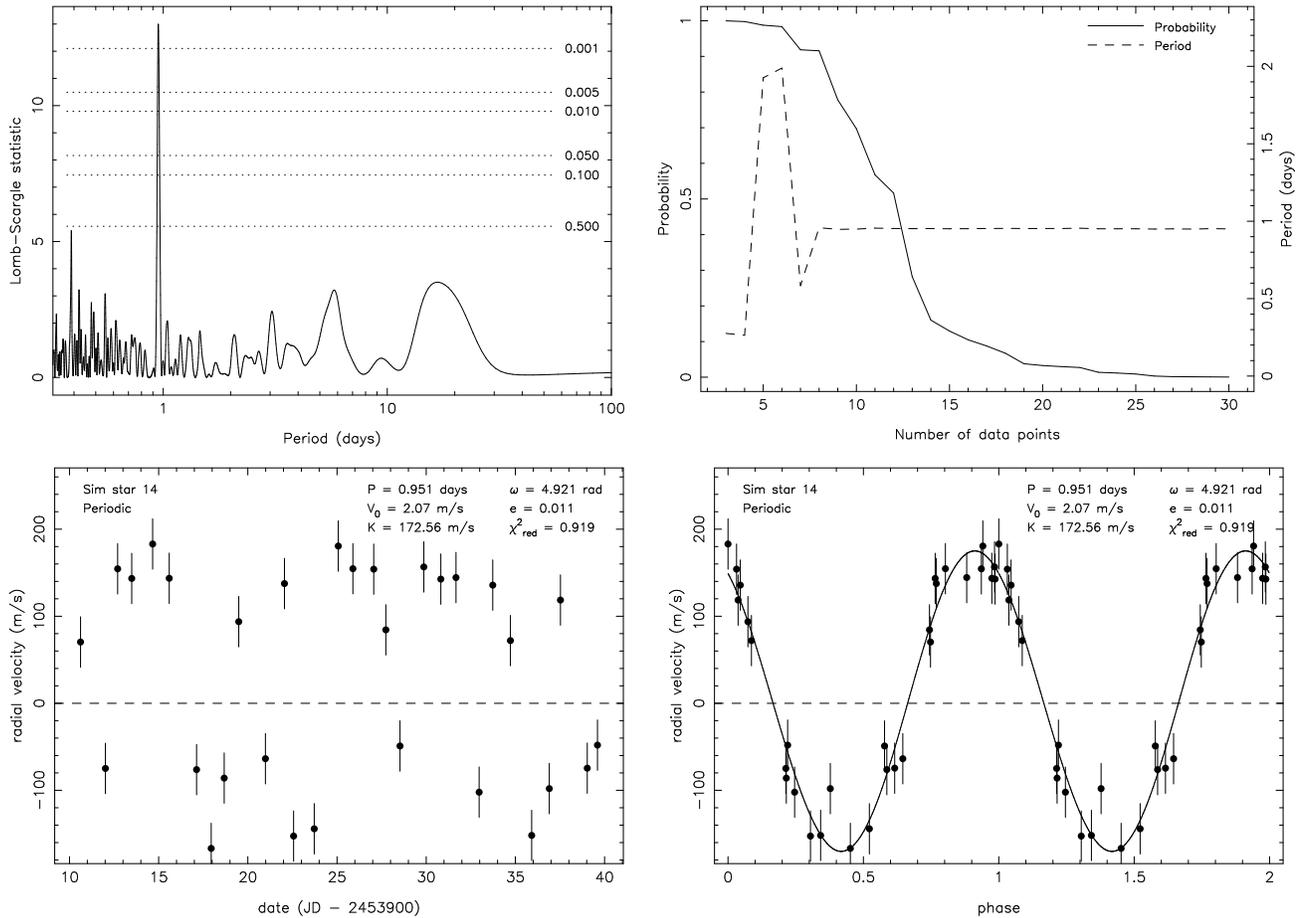

  \begin{center}
    \begin{tabular}{cc}
      \includegraphics[angle=270,width=8.2cm]{figure07a.ps} &
      \includegraphics[angle=270,width=8.5cm]{figure07b.ps} \\
      \includegraphics[angle=270,width=8.2cm]{figure07c.ps} &
      \includegraphics[angle=270,width=8.2cm]{figure07d.ps} \\
    \end{tabular}
  \end{center}
  \caption{An example of a $\sim 1$ day period planet observed for 30
    consecutive nights. The time of observation each night is randomized by
    passing identical times through a 4 hour gaussian filter. The resulting
    periodic signal in this case is easily recovered, as shown by the
    periodogram (top-left). The probability of the strongest period being a
    false-alarm drops significantly beyond $\sim 8$ data points (top-right).
    The unfolded (bottom-left) and folded data with best-fit model
    (bottom-right) are also shown.}
\end{figure*}

There are a variety of factors which can possibly contribute to an overall
bias of radial velocity surveys against detecting 1 day or even 2 day period
planets. Transit surveys are of course very strongly biased towards detecting
shorter period planets due to the geometric transit probability being so much
higher in this regime. In fact there is also a slight bias towards detecting
planets via the transit method around small (late-type) stars since the
depth of the transit lightcurve is significantly larger. This bias has thus
far been balanced by the difficulties imposed in achieving sufficient
photometric precision of M dwarfs and also that magnitude-limited wide-field
transit surveys tend to be dominated by a similar distribution of F--G--K
stars as those surveyed by radial velocity experiments. The smaller Roche
limit for M dwarfs may also be useful in explaining the difference in period
distributions of the transit and radial velocity surveys though, as pointed
out by \citet{for06}, the current small number statistics of the known
transiting planets limits such an investigation.

One of the main contributing factors to the radial velocity bias is undoubtedly
the adopted observing strategy of the survey in question. A typical survey
strategy will obtain one data point per target per night. In addition, it is
also typical to use the same observing schedule each night, usually to secure
observations when the target is closest to the zenith and hence at minimum
airmass. In the case where several observations are obtained for a single
target on a single night, these are often binned together to produce a single
high S/N image. This results in effectively observing the target at
approximately the same time every night.

In principle, observing a target once per day does not immediately exclude 1
day period planets from detection. Figure 7 shows an example of a case in
which a planetary signature with an approximately 1 day period is easily
recovered relatively quickly during 30 consecutive nights of observations. The
times of observation for the simulated data were produced by taking an identical
time of observation for each night and then passing each time through a gaussian
filter with a standard deviation of 4 hours. This method of randomizing the time
of observation results in a large variation of observing times from night to
night. The top-right plot in Figure 7 shows the dramatic drop in false-alarm
probability on a significant period estimate beyond $\sim 8$ data points, at
which point the estimate has settled upon the correct 1 day period.

The particular aspect of the observing strategy with the most impact is the
change in observing time per target per night. A quantitative estimate of this
impact was calculated by performing a Monte-Carlo simulation which produced
several thousand simulated radial velocity datasets containing planetary
signatures with periods between 0.9 and 1.1 days. The datasets assumed 30
consecutive nights and the time of observation each night was passed through
a gaussian filter with a standard deviation ranging from 1 to 4 hours. A
weighted Lomb-Scargle periodogram was calculated for each dataset and the
difference in false-alarm probability between the two highest peaks was used
as a quality estimate into the ambiguity of the period. Through this
simulation, an assessment of the effect of observing time on the likelihood
of planet detection was produced.

Figure 8 shows the results of the simulation. As demonstrated by Figures 7 and
8, the chances of being able to reliably determine the period of a short period
planet are exceptionally high when times of observations are sufficiently
scattered between nights. However, when the times of observations are quite
close, within an hour of each other for example, then the periodogram begins
to be plagued with aliases. Thus, increasing the time range over which data is
obtained for a particular target greatly reduces the relative strength of
contaminating period aliases. It is often the case that 1 day period aliases
are discarded from radial velocity surveys due to their ambiguous nature. The
signature of planets with periods close to 1 day are then particularly
vulnerable to the effect shown in Figure 8. Transit surveys also suffer
contamination by 1 day period aliases, though this is due to factors such as
night-to-night variations and red-noise rather than the sampling intervals.

\begin{figure}
  \includegraphics[angle=270,width=8.2cm]{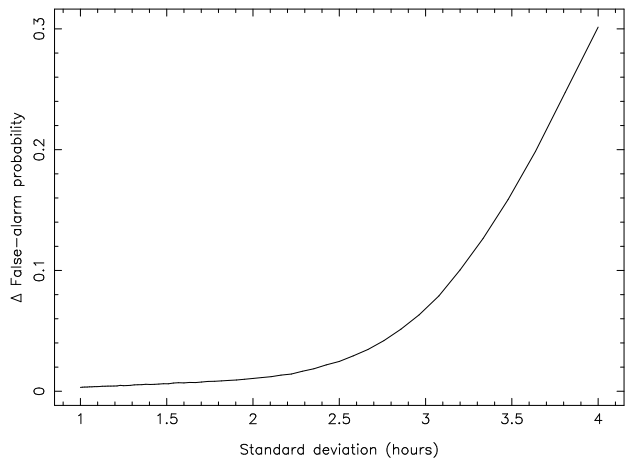}
  \caption{The change in the difference between the two highest periodogram
    peaks as a function of the standard devation in the time of observation.
    The quality of the period determination steadily increases when the target
    is observed at significantly different times each night.}
\end{figure}

There is a chance also of missing two day period planets if observations only
occur (either due to weather or other interuptions) every other day. The effect
is compounded by the fact that most radial velocity surveys only observe a
target long enough to ascertain if the target is variable. This is typically
assessed after a handful (5--10) of data points have been acquired. The effect
described here certainly does not completely account for the lack of very hot
Jupiters detected via radial velocity surveys. By providing a quantitative
estimate of the effect however, modifying observing strategies accordingly may
improve sensitivity to those planets in the future. Solutions to the problem
depend upon the times at which the target is observable and will usually
compromise observing time. An example is to observe the target at a similar
airmass either side of the meridian.

\section{Discussion}

The rate at which radial velocity planets are being discovered is increasing, a
trend which is expected to continue as next-generation radial velocity surveys
are realised. Considering the high number of expected hot Jupiters from current
and future surveys, the techniques presented here will be useful in planning
follow-up photometric observations of targets. Provided these follow-up
observations of the planet candidates are conducted within a reasonably short
time interval, the predicted transit windows will allow swift confirmation of
whether or not the planet's orbit is favourably inclined for transits to be
visible.

Surveys such as that being conducted by the N2K Consortium \citep{fis05} are
already employing such techniques, and has allowed them to make at least one
successful transit detection \citep{sat05}. The methods presented here are
part of an end-to-end system which allows the entire follow-up observing program
to be designed around the information available from the radial velocity
measurements and produced in a semi-automated fashion. The {\tt rvsim} code
mentioned here and described further in \citet{kan07} was used, for example, in
calculating the predicted transit ephemeris for HD 102195b \citep{ge06a}.
Follow-up observations were conducted during the predicted transit windows
using an Automatic Photoelectric Telescope \citep{hen99} located at Fairborn
Observatory. These observations rule out transits of the planetary companion.

The probability of an exoplanet transiting its parent star increases
dramatically with decreasing period. Perhaps ironically, the very hot Jupiters
with the highest geometric transit probability remain elusive to the radial
velocity surveys. This may be helped slightly by designing an observing
schedule which uses a large scatter in observing times on consecutive nights,
although this only applies to planets whose period are close to a small integer
number of days. Once radial velocity surveys begin to uncover these ultra-short
period planets, the number of transiting planets detected via the radial
velocity method will correspondingly increase. Even so, radial velocity planets
with periods close to an integer number of days are difficult to follow-up
photometrically because the transit phasing shifts very slowly with respect to
the night-day occurrence.

There are several multi-object radial velocity surveys for extra-solar planets
which are currently underway, using such facilities as GIRAFFE and FLAMES-UVES
on the VLT \citep{loe07} and a dispersed fixed-delay interferometer instrument,
called the Keck ET \citep{ge06b}, on the Sloan Digital Sky Survey 2.5m telescope
\citep{gun06}. The use of these instruments to survey such fields as the Kepler
field, as suggested by \citet{kan07}, will provide invaluable information for
the Kepler mission \citep*{bas05} regarding stars with companions; whether they
be planetary or stellar in nature. In particular, photometry of suspected
long-period planets for which only a handful of transits are observed will
benefit greatly from existing radial velocity data which are able to immediately
confirm their planetary nature. Considering that Kepler is expected to discover
planets with periods larger than 1 year during the mission lifetime, this is an
important factor to consider.

\section{Conclusions}

Simple calculations based upon the frequency of hot Jupiters and the
geometric transit probability show that around 0.1\% of solar-type stars
should have a detectable transit due to an extra-solar planet. In practise
however, the number of transiting planets detected via transit surveys is
much reduced, largely due to the issues of false positives and correlated
``red'' noise. This paper describes strategies for efficient photometric
follow-up of radial velocity planets to maximise the number of transit
discoveries. Since the target star is known to harbour a planet, the transit
detection is unaffected by many of the problems posed by blind transit
surveys and is simply limited by the transit probability.

The radial velocity and transit techniques of exoplanet detection are
complimentary in the information they provide regarding orbital and physical
parameters. For longer period planets, there is an increasing chance of
orbital eccentricity which can significantly alter the predicted time of
transit. Though the transit probability decreases in this regime, the
expected increase in planet discoveries at longer orbital periods will lead
to transiting planets found in eccentric orbits. The transit window must
also be handled with care, in particular taking into account the increase in
window size due to the period uncertainty. The fortran code {\tt rvsim} is
designed to automatically ingest, test for periodocity, and fit models to
large amounts radial velocity data from planet surveys. In addition, a transit
ephemeris is computed together which the observing windows, and the required
S/N for successful detection is estimated. In this way, an end-to-end process
has been constructed which is able to produce a customised observing campaign
complete with the telescope and instrument requirements. We present an
example using a 1.0m telescope, the class of which is relatively available
and make an ideal choice for fast follow-up of stars with known radial
velocity planets.

The discovery of transiting planets can be increased with improved observing
strategies for radial velocity surveys which eliminate any biases against
1--2 day period planets. We have demonstrated a possible cause for such a
bias and have suggested a programmed observing schedule which increases the
standard deviation in the times of observation. The ultimate goal is to
facilitate a radial velocity survey design in such a way that they are more
efficient at detecting very hot Jupiters which of course have the highest
chance to transit their parent star.

Although the detection of transiting planets via radial velocity surveys is
largely free from the observational biases which plague ``blind'' transit
surveys, the true distribution of transiting planets probably won't be fully
realised until missions such as Kepler begin their much anticipated
observations. These surveys are not without their own biases, but the large
expected number of transiting planets detected will yield unprecedented
insight into the formation and properties of these short-period planets.

\section*{Acknowledgements}

The author would like to thank Suvrath Mahadevan, Julian van Eyken, Scott
Fleming, and Robert Wilson for several useful discussions. The author is also
grateful for the constructive suggestions of the referee.

\end{document}